\documentstyle[twocolumn,aps]{revtex}
\input{epsf}

\begin{document}
\draft
\title{Properties of low--lying states in~a~diffusive~quantum~dot and
  Fock--space localization}

\author
{Carlos Mej{\'\i}a-Monasterio$^1$, Jean Richert$^2$, Thomas Rupp$^3$
  and Hans A. Weidenm{\"u}ller$^3$}
\address{
$^1$Instituto de F{\'\i}sica, Lab. de Cuernavaca, UNAM Apdo. postal
139-B, 62191, Cuernavaca, M\'exico}
\address{
$^2$Laboratoire de Physique Th\'eorique, ULP, 3 rue de l'Universit\'e,
67084 Strasbourg, France}
\address{
$^3$Max-Planck-Institut f{\"u}r Kernphysik, Postfach 103980, 69029
Heidelberg, Germany}
 
\date{\today}
 
\maketitle

\begin{abstract}Motivated by an experiment by Sivan {\it et al.}
(Europhys. Lett. 25, 605 (1994)) and by subsequent theoretical work on
localization in Fock space, we study numerically a hierarchical model
for a finite many--body system of Fermions moving in a disordered
potential and coupled by a two--body interaction. We focus attention
on the low--lying states close to the Fermi energy. Both the spreading
width and the participation number depend smoothly on excitation
energy. This behavior is in keeping with naive expectations and does
not display Anderson localization. We show that the model reproduces
essential features of the experiment by Sivan {\it et al.}

\end{abstract}



The measurement of the quasiparticle spectrum of a diffusive quantum
dot via its tunneling conductance by Sivan \emph{et~al.} \cite{sivan1}
in 1994 has caused considerable theoretical activity. The experimental
spectrum displayed a few narrow peaks in the vicinity of the Fermi
energy, followed by a quasi--continuum~\cite{sivan2}. The number of
discrete peaks was found to be of the order of the dimensionless
conductance $g$ of the dot. The results have given rise to a debate on
the validity of Fermi liquid theory for diffusive quantum dots. In
this theory, the low--lying excitations of a system of interacting
Fermions are described as quasiparticles, i.e., free Fermions with a
renormalized mass and a finite life time $\tau$, visible as peaks of
width $\hbar/\tau$ in the spectrum. Are these predictions consistent
with the very limited number of peaks visible in the spectrum of
Ref.~\cite{sivan1}, or does disorder invalidate Fermi liquid theory?

Altshuler \emph{et~al.}~\cite{agkl} asked (and answered) this question
in a precise way. These authors related the many--body problem with
electron--electron interaction in a diffusive quantum dot to that of
single--electron Anderson localization in real space. Many--body Fock
states are introduced as Slater determinants of eigenstates of the
single--particle Hamiltonian containing kinetic energy and disorder
potential. The distance in Fock space between two such states is
defined as twice the minimal number of electrons which have to be
moved from one single--particle state to another in order to get from
one state to the other. Matrix elements of a two--body interaction
between two states differ from zero only if the distance between the
two states is $\leq 4$. After introducing the Fermi energy and a
particle--hole representation, the states are grouped into classes,
each class being defined by the number of particle--hole ($p - h$)
excitations. The distance between classes is defined as the minimum
distance between any state in one class and any state in the other. In
the work of Ref.~\cite{agkl}, only couplings with distance 2 were
considered. Moreover, only terms that increase the complexity of the
states were kept, i.e., couplings to all states in the same and to all
but one state in the next--lower class were neglected. With these
assumptions, the Fock--space problem could be mapped onto a
tight--binding model on the infinite Bethe lattice.

For $g \gg 1$, the existence of three regimes separated by two
characteristic energies was shown. The Anderson
transition~\cite{anderson} between localized and extended states
occurs at the energy 
$E^{**}\!\sim\!\Delta \sqrt{g/\ln{g}}$, while 
$E^{*}\!\sim\!\Delta \sqrt{g}$ defines an effective energy above
which the many--particle states are completely mixed. Here, $\Delta$
is the average single--particle level spacing. These results were
corroborated by a calculation using supersymmetry \cite{mirlin} and
later also discussed in relation to the problem of few interacting
particles in a random potential~\cite{imry}, to the two--body random
interaction model~\cite{jacquod}, and to the level statistics of
excited many--body states~\cite{berkovits}. In Ref.~\cite{silvestrov},
doubts were voiced on the claim for a delocalization transition. In
Ref.~\cite{mirlin}, finite--size effects not considered in the
calculation using the Bethe lattice were estimated to yield
$E_{\rm ch}\!\sim\!\Delta g^{2/3}$ for the value of the energy at
which the states become completely mixed.

The work of Refs.~\cite{agkl,mirlin} uses a number of approximations
which are needed to establish the connection with the Bethe
lattice. In addition, it is assumed that the Bethe lattice has {\it
  infinite} length. These approximations are presumably valid at
sufficiently high excitation energy $E$. On the other hand, the data
of Ref.~\cite{sivan1} refer to the immediate vicinity of the Fermi
energy. For the interpretation of these data, it is important to know
whether the results of the high--energy approximations apply. For
instance, at low excitation energy and with $\varepsilon = E/\Delta$,
the number of accessible classes has a strong cutoff
$\sim{\sqrt{\varepsilon}}$.

To answer this question, we investigate in this Letter the low--lying
states of the many--body problem with disorder and interaction in the
framework of a model which is more realistic than but retains the
spirit of Refs.~\cite{agkl,mirlin}. We avoid the
approximations made in these papers which are needed to obtain the
structure of the Bethe lattice. We pay the price that we cannot use
analytical approximations and must rely on numerical simulation.


The Hamiltonian for spinless electrons (Fermions) has the form $H\!=\!
H_0\!+\!V$, with $H_0\!=\!\sum_k \epsilon_k a_k^{\dagger} a_k^{}$ the
unperturbed Hamiltonian, a sum of single--particle operators, and 
$V$ the two--body interaction. The symbol $\epsilon_k$ denotes the
single--particle eigenvalues, and $a_k^{\dagger}$ generates the
single--particle states $|k\rangle$. The single--particle Hamiltonian
contains a random potential. Therefore, the states $|k\rangle$ and
energies $\epsilon_k$ have stochastic properties. In an energy
interval of length $g \Delta$, these properties coincide with those of
the Gaussian orthogonal ensemble (GOE) of random matrices. We use the
classification scheme defined above and consider classes $m$ of Fock
states with $m$ particles and $(m\!-\!1)$ holes. Here, $m\!=\!1,2,\ldots.$ 
We denote the states by $|m,i\rangle$, where $i$ is an index running
over the states of the class. The associated energies, given by sums
of the $\epsilon_k$'s, are denoted by $E_{m,i}$, so that
$H_0|m,i\rangle\!=\!E_{m,i}|m,i\rangle$. To implement this model, we take
the energies $\epsilon_k$ after unfolding from the center of the
Wigner semicircle for the GOE. We choose the Fermi energy equal to
zero.

The unperturbed mean level density $\rho_m^{0}(\varepsilon)$ in class
$m$ is given by \cite{mirlin} $\rho_m^{0}(\varepsilon) = \Delta^{-1}
\varepsilon^{2m} / [(m+1)!m!(2m)!]$, where $\varepsilon = E/\Delta$
and $E$ is the excitation energy. With $\varepsilon$ fixed,
$\rho_m^{0}$ grows strongly with $m$ until it suddenly drops to almost
zero at $m \sim \sqrt{\varepsilon}$. Hence, at any value of
$\varepsilon$ only a limited number of classes contributes to the
total level density of the unperturbed system ($V = 0$). In the
vicinity of the Fermi energy, this number is one or two. This fact is
important for the experiment of Ref.~\cite{sivan1}. Indeed, if the
localization length is larger than two, then localization in Fock
space can have no bearing on the spectrum in the vicinity of the Fermi
surface.

The interaction operator $V$ mixes the states $|m,i\rangle$. We
suppose that the diagonal part of the interaction is included in $H_0$
by use of the Hartree--Fock method, without affecting the statistical
properties of either the energies $E_{m,i}$ or of the states
$|m,i\rangle$. The matrix elements of $V$ between two different Fock
states $|m,i\rangle$ and $|n,j\rangle$ vanish unless $|m-n|=0$~or~$1$
and unless the Fock distance of both states is $\leq 4$. (The case
$|m\!-\!n|=2$ 
requires creation or destruction of two particle--hole pairs out of
the Fermi sea. Such processes contribute to the unlinked diagrams of
perturbation theory and are not considered here. Calculations including
couplings with $|m\!-\!n|=2$ yielded results which did not differ
significantly.)
The non--vanishing matrix elements of $V$ are assumed~\cite{agkl,blanter}
to have a Gaussian distribution centered at zero with 
variance 
$\overline{V^2}=(\Delta/g)^2$. The mixing of the states
$|m,i\rangle$ depends on their spacing and on the strength of $V$,
i.e., on the value of $g$. In our model, the strength of $V$ does not
depend on excitation energy. This fact will cause stronger mixing of
the higher--lying states with smaller spacings. The effect is
compensated because out of the larger number of states, a decreasing
fraction couples to a given one. We allow for couplings between states
in the same class. This introduces terms which are explicitly excluded
in the case of the Bethe lattice.

In our numerical work, the energy scale is defined by $\Delta$. Our
only parameter is the dimensionless conductance $g$. The
dimension of the matrices was limited by a cutoff
$E_{\mathrm{cut\,off}}\!\gg\!\Delta$: All Fock states with energies
$E_{m,i}\!\geq\!E_{\mathrm{cut\,off}}$ were omitted. For each
realization of $H$, $E_{\mathrm{cut\,off}}$ was chosen
in such a way that the matrix dimension was 1000. This corresponds
roughly to $E_{\mathrm{cut\,off}}\!\sim 18\Delta$. Since our
single--particle energies are drawn from the center of a GOE
distribution, consistency requires that the Thouless energy $g \Delta$
is at least as large as the energy interval considered. The
eigenstates $|\alpha\rangle$ and eigenvalues $E_{\alpha}$ of the full
problem were obtained by diagonalization. We have checked that the
results presented below are independent of $E_{\mathrm{cut\,off}}$.
We now present two statistical measures suitable for a test of
localization in Fock space. 


\begin{figure}[t]
\vspace{-0.1cm}
\centerline{\epsfxsize=0.95\columnwidth \epsffile{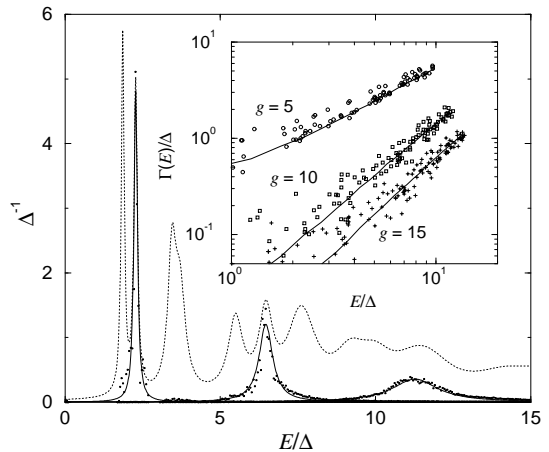}}
\caption{Quasiparticle line shapes $\rho_{1,i}$ for the $1p\!-\!0h$
  states $|1,2\rangle$, $|1,6\rangle$, and $|1,11\rangle$ (full
  curves). The dotted curve is the quasiparticle spectrum of
  all $1p\!-\!0h$ states. The inset shows the spreading width $\Gamma$ 
  versus $E$ for $g=5$ (circles), $g=10$ (squares), and $g=15$
  (crosses) together with the Golden Rule predictions $\Gamma_{\rm
  GR}(E)$ (full curves).}
\label{fig_qpspectrum}
\vspace{0.2cm}
\end{figure} 

\emph{Quasiparticle spectrum.} Adding an electron to a quantum dot with a
filled Fermi sea creates a $1p\!-\!0h$ Fock state $|1,i\rangle$. The
interaction $V$ spreads this state over a set of eigenstates
$|\alpha\rangle$. The spectral shape of the resulting quasiparticle
peak, also referred to as the local density of states, is given by
\begin{equation}
\rho_{1,i}(E) = \left\langle \sum_\alpha|\langle\alpha|1,i\rangle|^2\
\delta(E-E_\alpha) \right\rangle. 
\end{equation}
The brackets denote the ensemble average. In order to obtain
meaningful plots, we have used only a {\it single} realization of the
single--particle energies but $100$ realizations of the interaction
matrix elements. To reduce the density of the resulting data set, we
have averaged our results over small energy windows. This yields the
dots shown in Fig.~\ref{fig_qpspectrum} for three selected $1p\!-\!0h$
states at $g=10$. The fits with the Lorentzian 
\begin{equation}
  \frac{1}{2\pi}\frac{\Gamma_{1,i}}{(E-\epsilon_{1,i})^2 +
  \Gamma_{1,i}^2/4}
\end{equation}
(full curves) fix the centers $\epsilon_{1,i}$ of the quasiparticle
peaks and the spreading widths $\Gamma_{1,i}$. Adding the Lorentzian
line shapes of all $1p\!-\!0h$ states yields the quasiparticle spectrum
shown as dotted line.

\begin{figure}[t]
\vspace{-0.1cm}
\centerline{\epsfxsize=0.95\columnwidth \epsffile{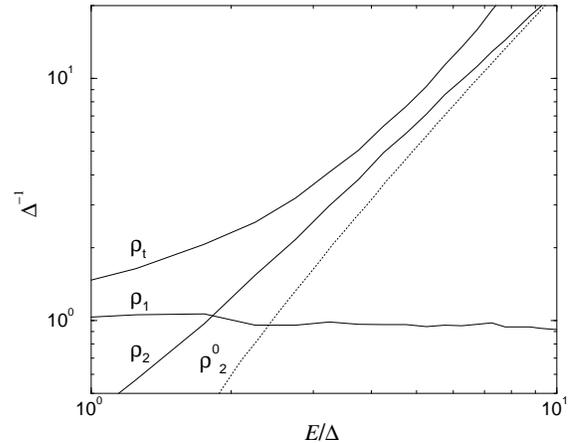}}
\caption{Densities $\rho_1(E)$, $\rho_2(E)$, and $\rho_t(E)$ for
  $g=10$ as defined in the text. For comparison we also show
  $\rho_2^{0}(E)$.}
\label{fig_densities}
\end{figure}

\emph{Spreading width.} Plotting $\Gamma_{1,i}$ for all states
$|1,i\rangle$ versus the state energies $\epsilon_{1,i}$, we obtain
the spreading width $\Gamma(E)$ as a function of excitation energy. 
For several values of $g$, this dependence is displayed in the inset
of Fig.~\ref{fig_qpspectrum}. We compare our results with the
prediction of the Golden Rule. We study the mixing of $1p\!-\!0h$
states. Each one of these is directly coupled to both all other
$1p\!-\!0h$ states and all $2p\!-\!1h$ states. The $1p\!-\!0h$
states have mean spacing $\Delta$. At the coupling strengths
($g\!\geq\!5$) shown in Fig.~\ref{fig_qpspectrum}, these states are
barely mixed with each other. Therefore we keep only the density of
$2p\!-\!1h$ states, $\rho_2(E)$, as the appropriate quantity to use in
the Golden Rule expression 
\begin{equation}
  \Gamma_{\rm GR}(E) = 2\pi \ (\Delta/g)^2 \ \rho_2(E).
\label{eq_goldenrule}
\end{equation}
We point out that this relation is valid beyond perturbation theory if
the density of final states is taken to be the exact (rather than the 
unperturbed) level density. The exact density in class $m$ is defined
as
\begin{equation}
  \rho_m(E)=\left\langle\sum_{i,\alpha}|\langle\alpha|m,i\rangle|^2\ 
  \delta(E-E_{\alpha})\right\rangle.
\label{dens}
\end{equation}
The brackets again denote the ensemble average which in practice we
perform over 100 realizations of the full Hamiltonian $H$. The three
full lines in the inset of Fig.~\ref{fig_qpspectrum} show $\Gamma_{\rm
  GR}(E)$ for the three values of $g$. With decreasing $g$, i.e.,
increasing interaction $V$, $\Gamma_{\rm GR}(E)$ increases while the slope
of $\Gamma_{\rm GR}(E)$ decreases. These facts reflect both the
$1/g^2$ dependence of $\overline{V^2}$ and the behavior of
$\rho_2(E)$. For $g = 10$, we show in Fig.~\ref{fig_densities}
$\rho_1(E)$, $\rho_2(E)$, $\rho_t(E) = \sum_m \rho_m(E)$, and the
unperturbed density of states $\rho_2^{0}(E)$. The latter differs 
substantially from $\rho_2(E)$: The interaction causes the density to
become wider, thereby reducing its slope. This accounts for the
behavior of $\Gamma_{\rm GR}(E)$ in the inset of 
Fig.~\ref{fig_qpspectrum}. In Refs.~\cite{agkl,mirlin}, it was
emphasized that localization in Fock space invalidates the Golden
Rule. The quantitative agreement between our numerical results and the
expression (\ref{eq_goldenrule}) down to very small energies shown in
Fig.~\ref{fig_qpspectrum} is, thus, a very strong argument against
the occurrence of localization in the energy range investigated in
this paper. The weak mixing of the low--lying states is entirely
explained by the low density of states right above the Fermi energy
and is not due to additional restrictions in Fock space.

\begin{figure}[t]
\vspace{-0.1cm}
\centerline{\epsfxsize=0.95\columnwidth \epsffile{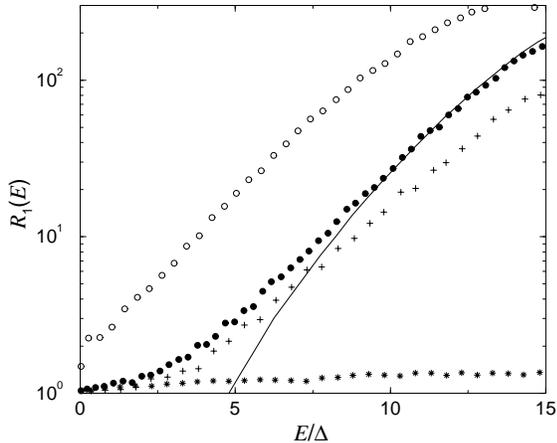}}
\caption{Participation number $R_{1}(E)$ of the $1p\!-\!0h$ states for
  $g=15$ (full circles) and $g=5$ (empty circles). The estimate
  $R_{1}^{\mathrm{GR}}(E)$ for $g=15$ is plotted as a full curve. In
  addition the results for applying a loopless Bethe lattice model
  (crosses) and considering couplings over Fock distance 2 only
  (stars) are shown (both $g=15$).}
\label{fig_participation}
\end{figure}

Qualitative features of the quasiparticle spectrum in
Fig.~\ref{fig_qpspectrum} are in good agreement with the results of
the experiment by Sivan~{\it et al.}~\cite{sivan1,sivan2}: A few
nearly discrete quasiparticle peaks with $\Gamma < \Delta$ occur right
above the Fermi energy. The width $\Gamma$ grows with excitation
energy until $\Gamma \approx \Delta$ at $E \approx g\Delta$ where the
spectrum becomes quasi--continuous. For an accurate simulation of 
Sivan's experiment, however, the $1p\!-\!0h$ excitations should be
built upon the ground state of {\em interacting} particles rather than
upon the filled Fermi sea. Calculations using such an interacting
ground state reproduced the results shown in Figs.~\ref{fig_qpspectrum} 
and \ref{fig_densities}, except for $g=5$. Here the spreading widths
were at the upper end of the statistical fluctuations shown in
Fig.~\ref{fig_qpspectrum}, while the Golden Rule underestimates
$\Gamma$ by about a factor two. The clear hierarchical structure
needed to study localization is lost when we start from an interacting
ground state. Therefore, we kept the simple model outlined above.
\vspace{0.2cm}
\emph{Participation number.} The average participation number often
serves as a measure of localization. The parti\-cipation number for the
Fock state $|m,i\rangle$ is defined by
\begin{equation}\label{eq_parrat}
  R_{m,i} = {\Big(\sum_{\alpha} {{|\langle \alpha|m,i \rangle|^4}
    \Big)^{-1}}}. 
\end{equation}
If the state $|m,i\rangle$ is localized, $R_{m,i} \approx 1$,
while $R_{m,i}$ increases monotonically with increasing mixing. In
the thermodynamic limit, $R_{m,i}$ is unbounded. In the case of finite
matrix dimension $N$ and the GOE, there exists an upper bound $R_{\rm
  max} = (1/3) N$ for $R_{m,i}$. In Fig.~\ref{fig_participation} we
show the average participation number $R_1(E)$ of class 1. This
quantity is obtained by plotting the values of $R_{1,i}$ for all
available $i$ versus the corresponding energies $E_{1,i}$ for a number
of realizations and averaging the result over energy. For weak
interaction (here $g=15$) and low excitation energies, the Fock states
$|1,i\rangle$ are localized because the spreading widths of the
quasiparticle states are small compared to the mean level spacing. 
With increasing $V$ ($g\!=\!5$) the mixing becomes larger even at low
excitation energies. We only find smooth transitions to mixed states,
both with increasing excitation energy $E$ and with increasing
interaction $V$. From the absence of any discontinuity in $R_1(E)$, we
conclude that there is no evidence for a localization transition in
Fock space. For $g=15$, the delocalization thresholds predicted in
Refs.~\cite{agkl,mirlin} have the values $E^{**} \approx 2.34 \Delta$
and $E^*\approx 3.87 \Delta$. We find $R_1(E) \lesssim 2$ for $ E \leq
5$ which shows that at these energies, the mixing of Fock states is
rather weak. More generally, an estimate of the number of eigenstates
$|\alpha\rangle$ contributing to a Fock state $|1,i\rangle$ at
energy~$E$ and, thus, of the average participation number $R_1(E)$ is
given by $R_1^{\rm GR}(E) = \Gamma_{\rm GR}(E) \ \rho_t(E)$ (see
Eq.~(\ref{eq_goldenrule}) and Fig.~\ref{fig_densities}). We find this
relation to be correct for sufficiently high energies $E$ (see
Fig.~\ref{fig_participation}) while $R_1^{\rm GR}(E)$ overestimates
$R_1(E)$ at low energies since by definition, $R_1(E)~\geq~1$ whereas
the level density becomes very small. For a localization transition,
we would expect $R_1(E)$ to lie below the Golden Rule estimate. This
is not the case.


\vspace{0.2cm}

\emph{Sensitivity to coupling scheme.} How do our results change with
coupling scheme, i.e., with the omission of those interaction terms which
spoil the analogy with the Bethe lattice? In the coupling scheme applied
so far we allowed for couplings between states of Fock distances 2 or
4, in keeping with the assumption of a two--body interaction. The
stars in Fig.~\ref{fig_participation} show the average participation
number for $1p\!-\!0h$ Fock states when only couplings between states
of Fock distance 2 are taken into account. This restriction means that
only one particle is allowed to change its single--particle state in an
interaction and no other particle can compensate the energy difference
implied by this transition. In effect couplings between states close
in energy are neglected.

In order to test the Bethe lattice assumption, i.e., a Fock space
topology without any loops, we first assumed that every Fock state
$|m,i\rangle$ with $m\!\geq\!2$ is coupled to only one state in class
$m\!-\!1$. This state was taken to be the one closest in energy to
$|m,i\rangle$. Couplings within each class were neglected. Couplings
with Fock--space distance 4 were taken into account. This model
yielded a set of disconnected trees each starting from one $1p-0h$
state. As each Fock state belongs only to exactly one tree, the trees
were all of rather small size. The resulting participation numbers
were much smaller than those of the full calculation, especially at
high energies. This is unsatisfactory. We then considered a second
model. We coupled each $1p\!-\!0h$ state to all $2p\!-\!1h$ states and
kept the couplings between higher classes as previously. Now all
Fock states except the $1p\!-\!0h$ states participate in each
tree. The resulting average participation numbers (crosses in
Fig.~\ref{fig_participation}) agree rather well with the full
calculation.


In conclusion, we presented a numerical study of the eigenvalues
and eigenfunctions of a finite Fermionic many--body problem with
random single--particle energies and a random two--body
interaction. Our model avoids the simplifications of the Bethe
lattice. Matrix diagonalization with a cutoff yielded results 
restricted to the vicinity of the Fermi surface and 
insensitive to the choice of the cutoff. We found that it is
important to keep couplings between states with a distance 4 in Fock
space. Moreover, we showed that for sensible values of the
conductance~$g$, our model is able to reproduce essential features of
the experiment of Ref.~\cite{sivan1}. Calculated values of the
spreading width and the participation number indicate some degree of
localization, depending on interaction strength and excitation energy. 
However, in contrast to analytical predictions based on a high--energy
approximation and on the Bethe lattice (which claim a twofold
localization transition) as well as those of Ref.~\cite{imry}, we have
only found a smooth transition from almost localized to delocalized
states. This transition can be fully understood in terms of the
density of available $1p\!-\!0h$ and $2p\!-\!1h$ states and does not
display Anderson localization. Our results show that the behavior of a
finite Fermi system at low excitation energy and at zero temperature
differs profoundly from the thermodynamic limit.

\emph{Acknowledgements.} We thank Y. Imry, T. Seligman, R. Jalabert
and T. Wilke for stimulating discussions.

\end{document}